\documentclass[aps,pra,notitlepage,superscriptaddress,showpacs]{revtex4-1}
\usepackage[T1]{fontenc}
\usepackage{textcomp}
\usepackage{amsmath}
\usepackage{amssymb}
\usepackage{graphicx}
\usepackage{esint}
\begin{document}
\title{Complete energy conversion between light beams carrying orbital angular
momentum using coherent population trapping for a coherently driven
double-$\Lambda$ atom-light coupling scheme}
\author{Hamid Reza Hamedi}
\email{hamid.hamedi@tfai.vu.lt}

\affiliation{Institute of Theoretical Physics and Astronomy, Vilnius University,
Saul\.etekio 3, Vilnius LT-10257, Lithuania}
\author{Emmanuel Paspalakis}
\email{paspalak@upatras.gr }

\affiliation{Materials Science Department, School of Natural Sciences, University
of Patras, Patras 265 04, Greece}
\author{Giedrius \v{Z}labys}
\email{giedrius.zlabys@tfai.vu.lt}

\affiliation{Institute of Theoretical Physics and Astronomy, Vilnius University,
Saul\.etekio 3, Vilnius LT-10257, Lithuania}
\author{Gediminas Juzeli\=unas}
\email{gediminas.juzeliunas@tfai.vu.lt}

\affiliation{Institute of Theoretical Physics and Astronomy, Vilnius University,
Saul\.etekio 3, Vilnius LT-10257, Lithuania}
\author{Julius Ruseckas}
\email{julius.ruseckas@tfai.vu.lt}

\affiliation{Institute of Theoretical Physics and Astronomy, Vilnius University,
Saul\.etekio 3, Vilnius LT-10257, Lithuania}
\begin{abstract}
We propose a procedure to achieve a complete energy conversion between
laser pulses carrying orbital angular momentum (OAM) in a cloud of
cold atoms characterized by a double-$\Lambda$ atom-light coupling
scheme. A pair of resonant spatially dependent control fields prepare
atoms in a position-dependent coherent population trapping state,
while a pair of much weaker vortex probe beams propagate in the coherently
driven atomic medium. Using the adiabatic approximation we derive
the propagation equations for the probe beams. We consider a situation
where the second control field is absent at the entrance to the atomic
cloud and the first control field goes to zero at the end of the atomic
medium. In that case the incident vortex probe beam can transfer its
OAM to a generated probe beam. We show that the efficiency of such
an energy conversion approaches the unity under the adiabatic condition.
On the other hand, by using spatially independent profiles of the
control fields, the maximum conversion efficiency is only $1/2$.
\end{abstract}
\maketitle

\section{Introduction}

The interaction of coherent light with atomic systems allow observation
of several important and interesting quantum interference effects
such as coherent population trapping (CPT) \citep{Alzetta76,Gray78,Radmore82,HioePhysRevA1988,CPT},
electromagnetically induced transparency (EIT) \citep{Harris-EIT-1997,Lukin-RevModPhys-2003,Fleischhauer-RevModPhys-2005,E.PaspalakisPRA2002}
and stimulated Raman adiabatic passage (STIRAP) \citep{Kuklinski1989,BergmannRevModPhys1998,Ivanov2004,Kumar2016,Antti2019}.
These phenomena are based on the coherent preparation of atoms in
a so-called dark state which is immune against the loss of population
through the spontaneous emission. Besides their fundamental interest,
these coherent optical effects have a number of useful applications
in various areas, such as enhanced nonlinear optics \citep{Harris-PhysRevLett-1990,Deng-PhysRevA-1998,Wang-phys.rev.lett.2001,Kang-PhysRevLett-2003},
slow light \citep{Harris-EIT-1997,Fleischhauer-RevModPhys-2005,Lukin-RevModPhys-2003,Juzeliunas-PhysRevLett-2004,Ruseckas2011PRA},
optical switching \citep{Braje2003} and storage of quantum information
\citep{ZibrovPhysRevLett2002}. 

A light beam can carry orbital angular momentum (OAM) due to its helical
wave front. Such a light beam with a spiral phase $e^{il\phi}$ has
an optical OAM of $\hbar l$ \citep{Allen1999,Miles-physToday-2004,AllenPhysRevA.45.8185,Babiker2019},
where $\phi$ denotes the azimuthal angle with respect to the beam
axis and $l$ is the winding number representing a number of times
the beams makes the azimuthal $2\pi$ phase shifts. The phase singularity
at the beam core of the twisted beam renders its donut-shaped intensity
profile. A number of interesting effects appear when this type of
optical beams interact with the atomic systems \citep{Babiker-PhysRevLett1994,Lembessis-PhysRevA-2010,Lembessis-PhysRevA.89-2014,Chen-PhysRevA-2008,Ding-OL-2012,WalkerPhysRevLett2012,RadwellPhysRevLet2015,SharmaPhysRevA2017,Hamedi-PhysRevA-2018,Hamedi2018OE,Ruseckas-PhysRevA-2007,Ruseckas-PhysRevA.-2011,Ruseckas-PhysRevA.87-2013,Dutton-PhysRevLett-2004,Hamedi2019,Ruseckas2011,Bortman-Arbiv-PhysRevA2001}.
Among them optical vorticities of slow light \citep{Dutton-PhysRevLett-2004,Ruseckas2011,Hamedi-PhysRevA-2018,Wang-PhysRevA-2008,Ruseckas-PhysRevA.87-2013}
have caused a considerable interest, as the OAM brings a new degree
of freedom in manipulation of the optical information during the storage
and retrieval of the slow light \citep{Pugatch-PhysRevLet-2004,Moretti-PhysRevA-2009}. 

The previous studies on the interplay of optical vortices and atomic
structures deal with the EIT situation where the atoms are initially
in their ground states. A four-level atom-light coupling of the tripod-type
was suggested to transfer optical vortices between different frequencies
during the storage and retrieval of the probe field \citep{Ruseckas-PhysRevA.-2011,Ruseckas2011}.
It has been demonstrated that without switching off and on of the
control fields, transfer of optical vortices take place by applying
a pair of weaker probe fields in the closed loop double-$\Lambda$
\citep{Hamedi-PhysRevA-2018} or double-tripod \citep{Ruseckas-PhysRevA.87-2013}
schemes. The exchange of optical vortices in non-closed loop structures
has been recently shown to be possible under the condition of weak
atom-light interaction in coherently prepared atomic media \citep{Hamedi2019}.
Such a medium is known as the phaseonium \citep{EberlyPRL1996,Scully-PhysRevLett-1985,Fleischhauer-PhysRevA-1992,PaspalakisPhysRevA2002CPM,PaspalakisPhysRevA2002multi,Paspalakis2002OL,Kis2003}.

The present paper extends the previous studies to a complete vortex
conversion based on CPT by employing a double-$\Lambda$ coherently
driven system. The double-$\Lambda$ configuration has been widely
employed due to its important applications in coherent control of
pulse propagation characteristics \citep{Deng2005,Hoonsoo2011,EilamPhysRevA.732006,ChiuPhysRevA2014,Turnbull2013,Shpaisman2005,Chong2008,Moiseev2006,Xiwei2013,Kang2006,EITpra2004,LiuPRL2016,Liu-PhysRevA-2004,Korsunsky-pra1999,IteOE2016,Ite2018pra,Shpaisman2004pra,Bortman-Arbiv-PhysRevA2001}.
Here we consider the propagation of laser pulses carrying optical
vortex in the double-$\Lambda$ system prepared in a superposition
state (dark state) of the lower $\Lambda$ subsystem induced by a
pair of spatially dependent control fields. It is shown that in an
adiabatic basis a single vortex beam initially acting on one transition
of the upper $\Lambda$ system generates an extra laser beam with
the same vorticity as that of the incident vortex beam with a unit
conversion efficiency. On the other hand, if the control fields have
spatially independent profiles, maximum vortex conversion efficiency
is restricted by the coherence between lower levels and can reach
only $1/2$. The proposed method to exchange optical vortices and
the complete vortex conversion is similar to the STIRAP, but the process
is reversed. In the usual STIRAP the atomic states are transferred
by properly choosing the time dependence of the light fields, whereas
here the energy is transferred between light beams by properly choosing
the position dependence of the atomic states, created by the position
dependence of the control light fields.

\section{Formulation}

\begin{figure}
\includegraphics[width=0.4\columnwidth]{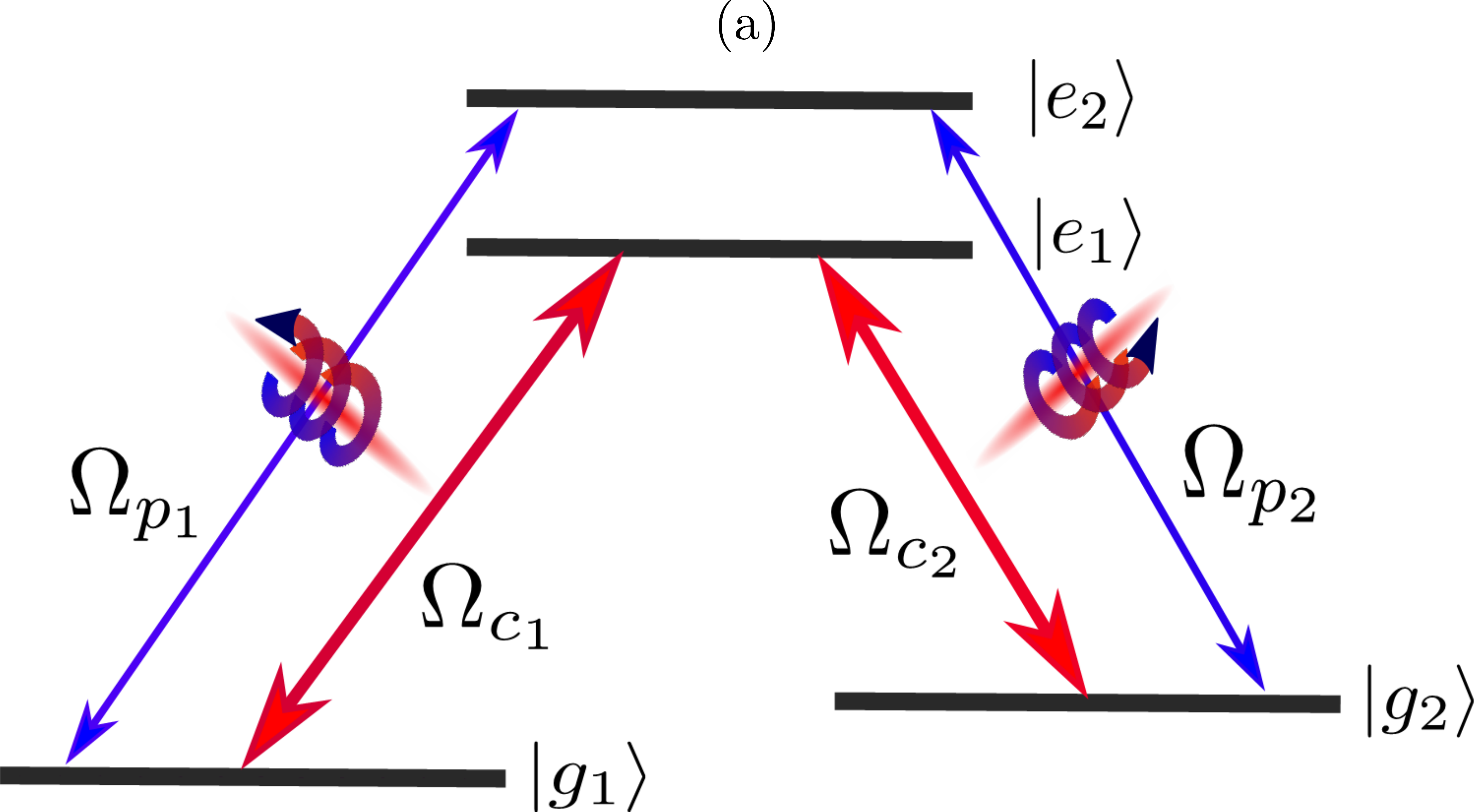} \includegraphics[width=0.5\columnwidth]{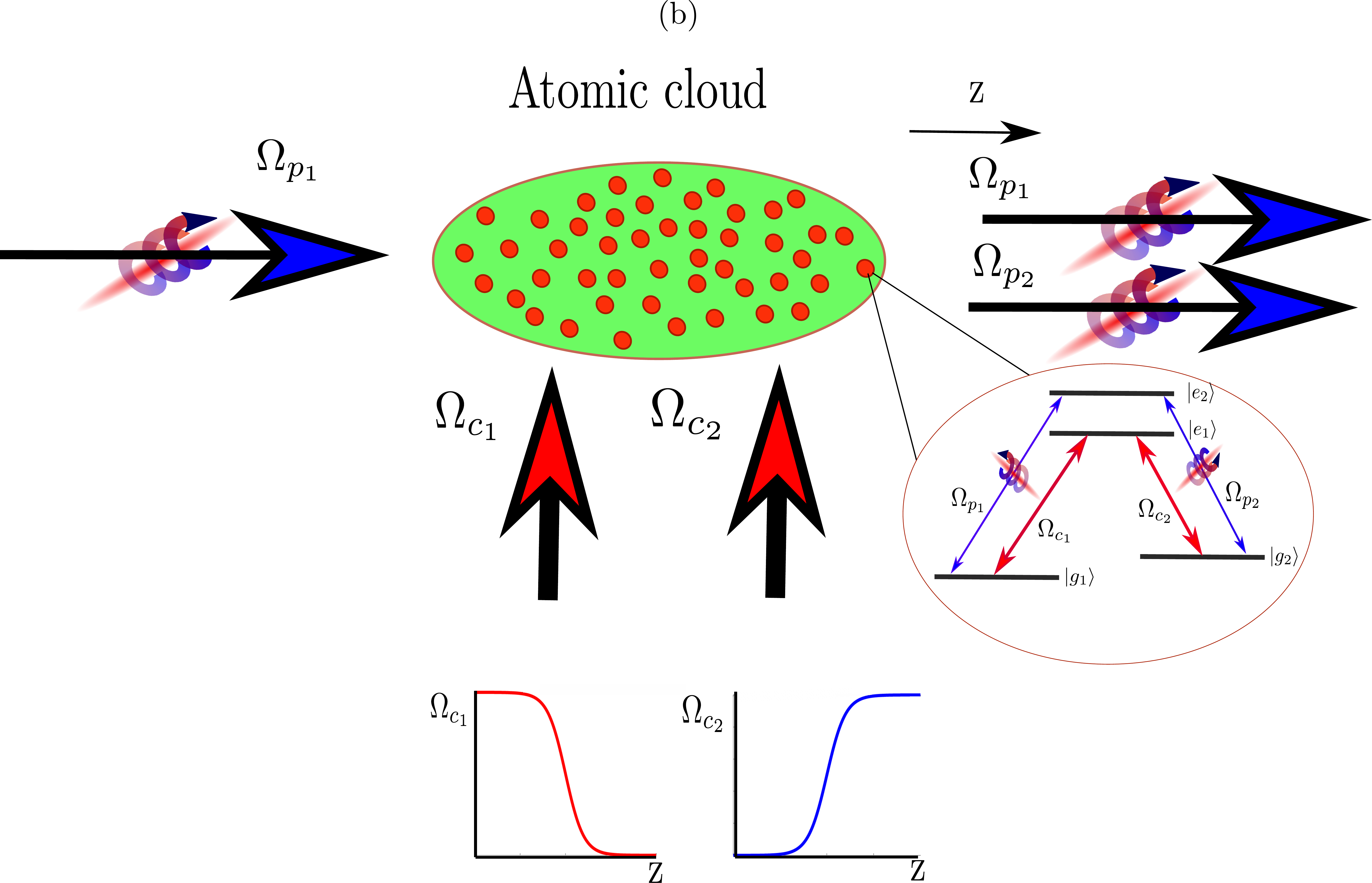}
\caption{(a) Schematic diagram of the double-$\Lambda$ atomic system. (b)
The schematic of possible arrangement of an experiment. Atoms (four-level
double-$\Lambda$ coupling scheme shown in red circle) inside atomic
cloud. A pair of resonant control fields $\Omega_{c_{1}}$ and $\Omega_{c_{2}}$(which
can be spatially dependent) prepare atoms in a CPT state (see down
bellow for spatial variation of control fields). A first vortex probe
beam $\Omega_{p_{1}}$, perpendicular to the control beams, propagates
inside the atomic cloud. The same vorticity of the first vortex beam
$\Omega_{p_{1}}$ is transferred to the second generated probe field
$\Omega_{p_{2}}$.}
\label{fig:1}
\end{figure}
Let us consider the double-$\Lambda$ scheme, depicted in Fig.~\ref{fig:1}
(a). We are interested in the propagation of two laser pulses with
the Rabi frequencies $\Omega_{p_{1}}$ and $\Omega_{p_{2}}$ in a
quantum medium consisting of atoms characterized by the double-$\Lambda$
configuration of the atom-light coupling. The two atomic lower states
$|g_{1}\rangle$ and $|g_{2}\rangle$ are coupled to two excited states
$|e_{1}\rangle$ and $|e_{2}\rangle$ via four laser fields characterized
by the slowly varying amplitudes $\Omega_{p_{1}}$, $\Omega_{p_{2}}$,
$\Omega_{c_{1}}$ and $\Omega_{c_{2}}$. 

Applying the rotating-wave approximation, the Hamiltonian for the
double-$\Lambda$ atomic system reads

\begin{equation}
H=-\Omega_{p_{1}}|e_{2}\rangle\langle g_{1}|-\Omega_{p_{2}}|e_{2}\rangle\langle g_{2}|-\Omega_{c_{1}}|e_{1}\rangle\langle g_{1}|-\Omega_{c_{2}}|e_{1}\rangle\langle g_{2}|+H.c.\,.\label{eq:hamiltoni}
\end{equation}
The Maxwell\textendash Bloch equations (MBE) governing the dynamics
of the probe fields $\Omega_{p_{1}}$ and $\Omega_{p_{2}}$ and two
atomic coherences $\rho_{e_{2}g_{1}}$ and $\rho_{e_{2}g_{2}}$ are
given by

\begin{equation}
\frac{\partial}{\partial t}\left[\begin{array}{c}
\rho_{e_{2}g_{1}}\\
\rho_{e_{2}g_{2}}
\end{array}\right]=-(i\delta+\Gamma)\left[\begin{array}{c}
\rho_{e_{2}g_{1}}\\
\rho_{e_{2}g_{2}}
\end{array}\right]+i\left[\begin{array}{cc}
\rho_{g_{1}g_{1}}-\rho_{e_{2}e_{2}} & \rho_{g_{2}g_{1}}\\
\rho_{g_{1}g_{2}} & \rho_{g_{2}g_{2}}-\rho_{e_{2}e_{2}}
\end{array}\right]\left[\begin{array}{c}
\Omega_{p_{1}}\\
\Omega_{p_{2}}
\end{array}\right]-i\rho_{e_{2}e_{1}}\left[\begin{array}{c}
\Omega_{c_{1}}\\
\Omega_{c_{2}}
\end{array}\right],\label{eq:OBE1}
\end{equation}
and

\begin{equation}
\frac{\partial}{\partial z}\left[\begin{array}{c}
\Omega_{p_{1}}\\
\Omega_{p_{2}}
\end{array}\right]+c^{-1}\frac{\partial}{\partial t}\left[\begin{array}{c}
\Omega_{p_{1}}\\
\Omega_{p_{2}}
\end{array}\right]=i\frac{\alpha\Gamma}{2L}\left[\begin{array}{c}
\rho_{e_{2}g_{1}}\\
\rho_{e_{2}g_{2}}
\end{array}\right].\label{eq:MaxwelEquation}
\end{equation}
Equation~(\ref{eq:OBE1}) for the atomic coherences is modeled by
means of the Liouville equation $\frac{\partial}{\partial t}\rho=-i[H,\rho]/\hbar+L_{\rho}$
with $L_{\rho}$ describing the decay of the system. On the other
hand, Eq.~(\ref{eq:MaxwelEquation}) for the evolution of the probe
fields is written under the slowly varying envelope approximation.
In the latter equation $\alpha$ determines the optical density of
the medium with a length $L$, and $\Gamma$ is the rate of spontaneous
decay rate of the excited states. We have assumed the same single
photon detuning for both probe fields in above equations, $\delta_{1}=\delta_{2}=\delta$,
where $\delta_{1}=\omega_{p_{1}}-\omega_{e_{2}g_{1}}$ and $\delta_{2}=\omega_{p_{2}}-\omega_{e_{2}g_{2}}$,
with $\omega_{p_{i}}$ being the central frequency of the corresponding
probe fields. The control fields are taken to be in exact resonance
with the corresponding atomic transitions.

The diffraction terms containing the transverse derivatives $(2k_{p_{1}})^{-1}\nabla_{\perp}^{2}\Omega_{p_{1}}$
and $(2k_{p_{2}})^{-1}\nabla_{\perp}^{2}\Omega_{p_{2}}$ have been
disregarded in the Maxwell equation (\ref{eq:MaxwelEquation}), where
$k_{p_{1}}=\omega_{p_{1}}/c$ and $k_{p_{2}}=\omega_{p_{2}}/c$ represent
the central wave vectors of the first and second probe beams. These
terms can be evaluated as $\nabla_{\perp}^{2}\Omega_{p_{1(2)}}\sim w^{-2}\Omega_{p_{1(2)}}$,
where $w$ is a characteristic transverse dimension of the laser beams.
It can be a width of the vortex core if the beam carries an optical
vortex or a characteristic width of the beam when there is no vortex.
The change of the phase of the probe beams due to the diffraction
term after passing the medium is then estimated to be $L/2kw^{2}$,
where $L$ is the length of the atomic cloud, with $k\approx k_{p_{1(2)}}$.
One can neglect the phase change $L/2kw^{2}$ when the sample length
$L$ is not too large, $L\lambda/w^{2}\ll\pi$, where $\lambda=2\pi/k$
is an optical wavelength. For example, by taking the length of the
atomic cloud to be $L=100\,\mathrm{\mu m}$, the characteristic transverse
dimension of the beams $w=20\,\mathrm{\mu m}$ and the wavelength
$\lambda=1\,\mathrm{\mu m}$, one obtains $L\lambda/w^{2}=0.25$.
Under these conditions the diffraction terms do not play an important
role and one can safely drop it out in Eq.~(\ref{eq:MaxwelEquation}).

Let us assume a situation where a pair of resonant strong control
fields acting on the lower legs of the double-$\Lambda$ prepare the
atoms in a CPT (or dark) state 

\begin{equation}
|D\rangle=\cos\theta_{c}|g_{1}\rangle-\sin\theta_{c}|g_{2}\rangle,\label{eq:atom-state1}
\end{equation}
where a mixing angle $\theta_{c}$ is defined by the equation

\begin{equation}
\left[\begin{array}{c}
\cos\theta_{c}\\
\sin\theta_{c}
\end{array}\right]=\frac{1}{\sqrt{|\Omega_{c_{1}}|^{2}+|\Omega_{c_{2}}|^{2}}}\left[\begin{array}{c}
\Omega_{c_{2}}\\
\Omega_{c_{1}}
\end{array}\right]\,.\label{eq:micangles}
\end{equation}
Consequently, a weak probe pulse pair propagates in a coherently prepared
medium interacting with the upper legs of double $\Lambda$ scheme.

Since the probe fields are much weaker than the control fields, $|\Omega_{p_{1}}|,|\Omega_{p_{2}}|\ll\sqrt{|\Omega_{c_{1}}|^{2}+|\Omega_{c_{2}}|^{2}}$,
the atomic coherence $\rho_{g_{2}g_{1}}$ as well as the populations
$\rho_{g_{1}g_{1}}$ and $\rho_{g_{2}g_{2}}$ change little during
the propagation of the probe fields, giving

\begin{equation}
\left[\begin{array}{c}
\rho_{g_{2}g_{1}}\\
\rho_{g_{1}g_{1}}\\
\rho_{g_{2}g_{2}}
\end{array}\right]=\left[\begin{array}{c}
-\sin\theta_{c}\cos\theta_{c}\\
\cos^{2}\theta_{c}\\
\sin^{2}\theta_{c}
\end{array}\right]\,.\label{eq:coherence-and-populations}
\end{equation}
In addition, when $|\Omega_{p_{1}}|,|\Omega_{p_{2}}|\ll\Gamma$, the
excited states are weakly populated, $\rho_{e_{2}g_{1}},\rho_{e_{2}g_{2}}\ll1$,
and Eq.~(\ref{eq:OBE1}) changes to 

\begin{equation}
\frac{\partial}{\partial t}\left[\begin{array}{c}
\rho_{e_{2}g_{1}}\\
\rho_{e_{2}g_{2}}
\end{array}\right]=-(i\delta+\Gamma)\left[\begin{array}{c}
\rho_{e_{2}g_{1}}\\
\rho_{e_{2}g_{2}}
\end{array}\right]+i\left[\begin{array}{cc}
\cos^{2}\theta_{c} & -\sin\theta_{c}\cos\theta_{c}\\
-\sin\theta_{c}\cos\theta_{c} & \sin^{2}\theta_{c}
\end{array}\right]\left[\begin{array}{c}
\Omega_{p_{1}}\\
\Omega_{p_{2}}
\end{array}\right],\label{eq:OBE2}
\end{equation}
Then, to the first order, the following steady state solutions for
the coherences $\rho_{e_{2}g_{1}}$, $\rho_{e_{2}g_{2}}$are obtained

\begin{equation}
(\delta-i\Gamma)\left[\begin{array}{c}
\rho_{e_{2}g_{1}}\\
\rho_{e_{2}g_{2}}
\end{array}\right]=\left[\begin{array}{cc}
\cos^{2}\theta_{c} & -\sin\theta_{c}\cos\theta_{c}\\
-\sin\theta_{c}\cos\theta_{c} & \sin^{2}\theta_{c}
\end{array}\right]\left[\begin{array}{c}
\Omega_{p_{1}}\\
\Omega_{p_{2}}
\end{array}\right]\,.\label{eq:solutions}
\end{equation}
Substituting Eq.~(\ref{eq:solutions}) into the Maxwell equations
(\ref{eq:MaxwelEquation}), one arrives at the following coupled equations
for the propagation of the pulse pair 

\begin{equation}
\frac{\partial}{\partial z}\left[\begin{array}{c}
\Omega_{p_{1}}\\
\Omega_{p_{2}}
\end{array}\right]=-iK\left[\begin{array}{c}
\Omega_{p_{1}}\\
\Omega_{p_{2}}
\end{array}\right],\label{eq:toWaveSolutions}
\end{equation}
with
\begin{equation}
K=\beta\left[\begin{array}{cc}
\cos^{2}\theta_{c} & -\sin\theta_{c}\cos\theta_{c}\\
-\sin\theta_{c}\cos\theta_{c} & \sin^{2}\theta_{c}
\end{array}\right],\label{eq:k}
\end{equation}
and

\begin{equation}
\beta=\frac{\alpha\Gamma}{2L(i\Gamma-\delta)}.\label{eq:10}
\end{equation}
From Eq.~(\ref{eq:toWaveSolutions}) follows that the probe field
intensity will oscillate when the two-photon detuning $\delta$ is
not zero \citep{Shpaisman2004pra,EilamPhysRevA.732006}. In addition,
as can be seen from Eqs.~(\ref{eq:toWaveSolutions}) and (\ref{eq:10}),
$\delta\neq0$ introduces additional probe filed losses because $\beta$
acquires a real part. In the following, we consider the case when
$\delta=0$ to minimize energy losses of the probe fields.

Let us assume that entrance to the medium the second probe field is
absent $\Omega_{p_{2}}(0)=0$, while the first probe field $\Omega_{p_{1}}$
is a vortex beam defined by 
\begin{equation}
\Omega_{p_{1}}(0)=\Omega(r)=\Omega_{p_{10}}\left(\frac{r}{w}\right)^{|l|}e^{-r^{2}/w^{2}}e^{il\phi},\label{eq:OP1}
\end{equation}
where $l$ is the orbital angular momenta along the propagation axis
$z$, and $\phi$ is the azimuthal angle, $r$ describes a cylindrical
radius, $w$ is a beam waist, and $\Omega_{p_{10}}$ represents the
strength of the vortex beam. The solutions to the Eq.~(\ref{eq:toWaveSolutions})
read

\begin{equation}
\left[\begin{array}{c}
\Omega_{p_{1}}(z)\\
\Omega_{p_{2}}(z)
\end{array}\right]=\Omega(r)\left[\begin{array}{c}
\cos^{2}\theta_{c}e^{-i\beta z}+\sin^{2}\theta_{c}\\
-\sin\theta_{c}\cos\theta_{c}\left(e^{-i\beta z}-1\right)
\end{array}\right].\label{eq:finalSolutions}
\end{equation}
Evidently, the laser beam $\Omega_{p_{1}}$ transfers its vortex to
the generated beam $\Omega_{p_{2}}$. The intensity of both new probe
field vortices can be manipulated by the Rabi-frequencies of control
fields (see Eq.~(\ref{eq:coherence-and-populations})).

\begin{figure}
\includegraphics[width=0.5\columnwidth]{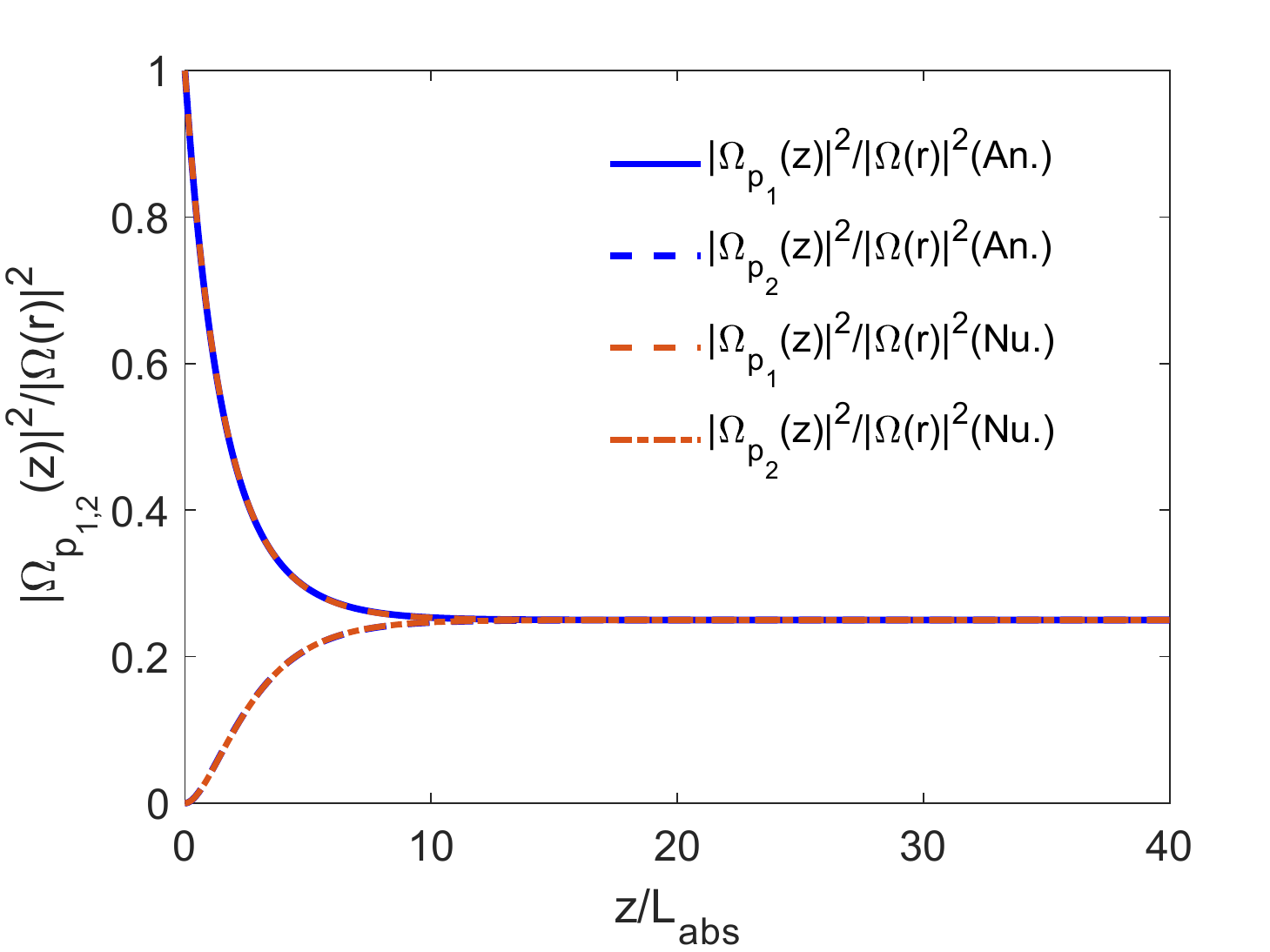} \caption{Analytical and numerical results of the dimensionless intensities
of the light fields $|\Omega_{p_{1}}(z)|^{2}/|\Omega(r)|^{2}$ and
$|\Omega_{p_{2}}(z)|^{2}/|\Omega(r)|^{2}$ against the dimensionless
distance $z/L_{abs}$ for $\Omega_{c_{1}}=\Omega_{c_{2}}=\Gamma$,
$\delta=0$ and $\alpha=40$. Analytical results are plotted based
on Eq.~(\ref{eq:finalSolutions}), while numerical results are based
on the full Optical-Bloch equations given by Eqs. (\ref{eq:A1})-(\ref{eq:A9})
together with the Maxwell equation (\ref{eq:MaxwelEquation}). For
the numerical simulations, the incident form of probe pulse is $\Omega_{p_{1}}(z=0,t)=\Omega_{p_{1}}^{0}e^{-(t-t_{0})^{2}/2\bar{t}^{2}}$while
$\Omega_{p_{2}}(z=0,t)=0$. The parameters used for the numerical
calculations are $\Omega_{p_{1}}^{0}=0.01\Gamma$, $t_{0}=25$ and
$\bar{t}=10$. Abbreviations \textquotedbl An.\textquotedbl{} and
\textquotedbl Nu.\textquotedbl{} in plots stand for Analytical and
Numerical results. }
\label{fig:2}
\end{figure}

At the beginning of the atomic cloud where the first probe vortex
beam $\Omega_{p_{1}}(0)$ has just entered, the second probe beam
is still absent ($\Omega_{p_{2}}(0)=0$), and it is generated going
deeper into the atomic medium. As can be seen from Eq.~(\ref{eq:finalSolutions}),
both vortex beams are not affected considerably by the losses during
their propagation as long as the optical density length is large enough.
In fact, the losses take place mostly at the entrance of the medium.
If optical density of the resonant medium is sufficiently large $\alpha\gg1$,
the absorption length $L_{\mathrm{abs}}=L/\alpha$ constitutes a fraction
of the whole medium $L_{\mathrm{abs}}\ll L$. For larger propagating
distances $z\gg L_{\mathrm{abs}}$, the losses disappear

\begin{equation}
\left[\begin{array}{c}
\Omega_{p_{1}}(z\gg L_{abs})\\
\Omega_{p_{2}}(z\gg L_{abs})
\end{array}\right]=\Omega(r)\left[\begin{array}{c}
\sin^{2}\theta_{c}\\
\sin\theta_{c}\cos\theta_{c}
\end{array}\right],\label{eq:longLimit}
\end{equation}
and the system goes to the dark state given by Eq.~(\ref{eq:atom-state1})
(see also Fig.~\ref{fig:2}).

Let us consider the efficiency of frequency conversion from initial
vortex beam $\Omega_{p_{1}}$ to the generated vortex beam $\Omega_{p_{2}}$
for $z\gg L_{\mathrm{abs}}$ described by Eq.~(\ref{eq:longLimit}).
The efficiency of vortex conversion between different frequencies
is

\begin{equation}
\eta=|\Omega_{p_{2}}(z\gg L_{abs})/\Omega_{p_{1}}(0)|=|\rho_{g_{1}g_{2}}|=\sin\theta_{c}\cos\theta_{c}.\label{eq:frequency conversion}
\end{equation}
Clearly, the maximum frequency conversion between vortex beams achieved
in this way is only $1/2$ (when the coherence $\rho_{g_{1}g_{2}}$
is maximum). In the next section, we present another scenario for
transfer of optical vortices in such a medium with the possibility
to achieve the unit efficiency of the vortex conversion.

\section{Spatially dependent control fields}

Let us consider a situation where two strong resonant control laser
fields $\Omega_{c_{1}}(z)$ and $\Omega_{c_{2}}(z)$ are spatially
dependent. This is the case if the control beams propagate perpendicular
to the probe beams (Fig.~\ref{fig:1} (b)), the latter beams propagating
along the $z$ axis. In that case $\Omega_{c_{1}}(z)$ and $\Omega_{c_{2}}(z)$
represent the transverse profiles of the control beams. Application
of such control fields prepare the system in the spatially dependent
dark state $|D(z)\rangle$ immune to the spontaneously decay 

\begin{equation}
|D(z)\rangle=\cos\theta(z)|g_{1}\rangle-\sin\theta(z)|g_{2}\rangle,\label{eq:position-dependent-dark-state}
\end{equation}
with

\begin{equation}
\left[\begin{array}{c}
\cos\theta(z)\\
\sin\theta(z)
\end{array}\right]=\frac{1}{\sqrt{|\Omega_{c_{1}}(z)|^{2}+|\Omega_{c_{2}}(z)|^{2}}}\left[\begin{array}{c}
\Omega_{c_{2}}(z)\\
\Omega_{c_{1}}(z)
\end{array}\right],\label{eq:mixangleZ}
\end{equation}
where $\theta(z)$ is the position-dependent mixing angle which differs
from the constant mixing angle $\theta_{c}$ used in previous section.
Again the interaction of probe fields with the medium is considered
to be sufficiently weak, so it does not affect significantly the evolution
of the density matrix, which is determined mostly by the coupling
fields entering the  density matrix equations. This means that one
can use the results from a strongly driven $\Lambda$ system for the
lower atomic states when the probe fields only act as a weak perturbation
that does not change the density matrix elements

\begin{equation}
\left[\begin{array}{c}
\rho_{g_{2}g_{1}}(z)\\
\rho_{g_{1}g_{1}}(z)\\
\rho_{g_{2}g_{2}}(z)
\end{array}\right]=\left[\begin{array}{c}
-\sin\theta(z)\cos\theta(z)\\
\cos^{2}\theta(z)\\
\sin^{2}\theta(z)
\end{array}\right].\label{eq:coherence-and-populations-Z}
\end{equation}
The approximate solutions to the density matrix equations read

\begin{equation}
(\delta-i\Gamma)\left[\begin{array}{c}
\rho_{e_{2}g_{1}}(z)\\
\rho_{e_{2}g_{2}}(z)
\end{array}\right]=\left[\begin{array}{cc}
\cos^{2}\theta(z) & -\sin\theta(z)\cos\theta(z)\\
-\sin\theta(z)\cos\theta(z) & \sin^{2}\theta(z)
\end{array}\right]\left[\begin{array}{c}
\Omega_{p_{1}}(z)\\
\Omega_{p_{2}}(z)
\end{array}\right],\label{OBS2}
\end{equation}
and the propagation equations become

\begin{equation}
\frac{\partial}{\partial z}\left[\begin{array}{c}
\Omega_{p_{1}}(z)\\
\Omega_{p_{2}}(z)
\end{array}\right]=-iK(z)\left[\begin{array}{c}
\Omega_{p_{1}}(z)\\
\Omega_{p_{2}}(z)
\end{array}\right],\label{eq:ME2}
\end{equation}
with 

\begin{equation}
K(z)=\beta\left[\begin{array}{cc}
\cos^{2}\theta(z) & -\sin\theta(z)\cos\theta(z)\\
-\sin\theta(z)\cos\theta(z) & \sin^{2}\theta(z)
\end{array}\right].\label{eq:Kz}
\end{equation}

There is no general analytical solution of the propagation equation
(\ref{eq:ME2}) as the coefficients of the propagation matrix Eq.~(\ref{eq:Kz})
are spatially dependent. Equation (\ref{eq:ME2}) is similar to a
time-dependent Schrodinger equation with the time-dependence replaced
with spatial dependence ($t\rightarrow z$). In this case, the propagation
matrix $K(z)$ featured in Eq.~(\ref{eq:Kz}) resembles a time-dependent
Hamiltonian. When the Hamiltonian is time-dependent, general solutions
are available if the dynamics satisfies adiabaticity \citep{BergmannRevModPhys1998,Paspalakis2002OL}.
Therefore, we follow a standard adiabatic approximation method to
study the evolution of the system. Yet the adiabatic evolution here
refers to the spatial evolution. The eigenstates of the propagation
matrix $K(z)$ in Eq.~(\ref{eq:Kz}) can be written as
\begin{align}
a^{0}= & \left[\begin{array}{c}
\sin\theta(z)\\
\cos\theta(z)
\end{array}\right],\label{eq:ES1}\\
a^{\beta}= & \left[\begin{array}{c}
\cos\theta(z)\\
-\sin\theta(z)
\end{array}\right],\label{eq:ES2}
\end{align}
where the two eigenstates $a^{0}$ and $a^{\beta}$ are orthogonal.
It is straightforward to verify that the corresponding eigenenergies
of the propagation matrix (\ref{eq:Kz}) are
\begin{align}
\lambda^{0}= & 0,\label{eq:EV1}\\
\lambda^{\beta}= & \beta.\label{eq:EV2}
\end{align}

Let us form a unitary matrix $U$ as

\begin{equation}
U(z)=\left[\begin{array}{cc}
\sin\theta(z) & \cos\theta(z)\\
\cos\theta(z) & -\sin\theta(z)
\end{array}\right].\label{eq:U}
\end{equation}
The matrix $U$ then transforms Eq.~(\ref{eq:ME2}) to the adiabatic
basis given by Eqs.~(\ref{eq:ES1})-(\ref{eq:EV2})

\begin{equation}
\frac{\partial}{\partial z}\left[\begin{array}{c}
\tilde{\Omega}_{p_{1}}(z)\\
\tilde{\Omega}_{p_{2}}(z)
\end{array}\right]=-i\tilde{K}(z)\left[\begin{array}{c}
\tilde{\Omega}_{p_{1}}(z)\\
\tilde{\Omega}_{p_{2}}(z)
\end{array}\right],\label{eq:WET}
\end{equation}
where

\begin{equation}
\left[\begin{array}{c}
\tilde{\Omega}_{p_{1}}(z)\\
\tilde{\Omega}_{p_{2}}(z)
\end{array}\right]=U(z)^{-1}\left[\begin{array}{c}
\Omega_{p_{1}}(z)\\
\Omega_{p_{2}}(z)
\end{array}\right],\label{eq:a}
\end{equation}
with the transformed propagation matrix $\tilde{K}(z)$ given by

\begin{equation}
\tilde{K}(z)=-iU(z)^{-1}\frac{\partial}{\partial z}U(z)+U(z)^{-1}K(z)U(z).\label{eq:knew1}
\end{equation}
In the matrix form it reads

\begin{equation}
\tilde{K}(z)=i\left[\begin{array}{cc}
0 & \cos\theta(z)\frac{\partial}{\partial z}\sin\theta(z)-\sin\theta(z)\frac{\partial}{\partial z}\cos\theta(z)\\
-\cos\theta(z)\frac{\partial}{\partial z}\sin\theta(z)+\sin\theta(z)\frac{\partial}{\partial z}\cos\theta(z) & \beta
\end{array}\right].\label{eq:knew2}
\end{equation}
As can be seen, the eigenvalue $\beta$ appears in the diagonal of
$\tilde{K}(z)$ which generally describes attenuation. However, for
sufficiently long propagation distances, the field component along
the eigenstate $a^{\beta}$ vanishes. The adiabaticity limit constrains
that the off-diagonal elements of $\tilde{K}(z)$ are negligible compared
to the difference of the diagonal ones

\begin{equation}
|\cos\theta(z)\frac{\partial}{\partial z}\sin\theta(z)-\sin\theta(z)\frac{\partial}{\partial z}\cos\theta(z)|\ll|\beta|.\label{eq:adiabaticity}
\end{equation}

When the adiabaticity condition (\ref{eq:adiabaticity}) is fulfilled,
the solution of the propagation equation (\ref{eq:ME2}) reads

\begin{equation}
\left[\begin{array}{c}
\Omega_{p_{1}}(z)\\
\Omega_{p_{2}}(z)
\end{array}\right]=U(z)WU(z)^{-1}\left[\begin{array}{c}
\Omega_{p_{1}}(z_{i})\\
\Omega_{p_{2}}(z_{i})
\end{array}\right],\label{eq:we2}
\end{equation}
where

\begin{equation}
W=\left[\begin{array}{cc}
e^{-i\int_{z_{i}}^{z}\tilde{K}_{11}(z)dz} & 0\\
0 & e^{-i\int_{z_{i}}^{z}\tilde{K}_{22}(z)dz}
\end{array}\right].\label{eq:W}
\end{equation}
 For a long propagation distance $(z\gg L_{abs})$, the exponential
term $e^{-i\int_{z_{i}}^{z}\tilde{K}_{22}(z)dz}$ vanishes. In this
limit the solutions to the Eq.~(\ref{eq:we2}) read

\begin{equation}
\left[\begin{array}{c}
\Omega_{p_{1}}(z)\\
\Omega_{p_{2}}(z)
\end{array}\right]=\Omega(r)\left[\begin{array}{c}
\sin\theta(z_{i})\sin\theta(z)\\
\sin\theta(z_{i})\cos\theta(z)
\end{array}\right],\label{eq:lastSolutions}
\end{equation}
where we have assumed that at the entry of the medium $\Omega_{p_{1}}(z_{i})$
is a vortex ($\Omega_{p_{1}}(z_{i})=\Omega(r)$) and $\Omega_{p_{2}}(z_{i})=0$.
As one can see from Eq.~(\ref{eq:lastSolutions}), in this case the
same vorticity of the first vortex beam $\Omega_{p_{1}}(z_{i})$ is
transferred to the second generated field $\Omega_{p_{2}}(z)$.

\begin{figure}
\includegraphics[width=0.4\columnwidth]{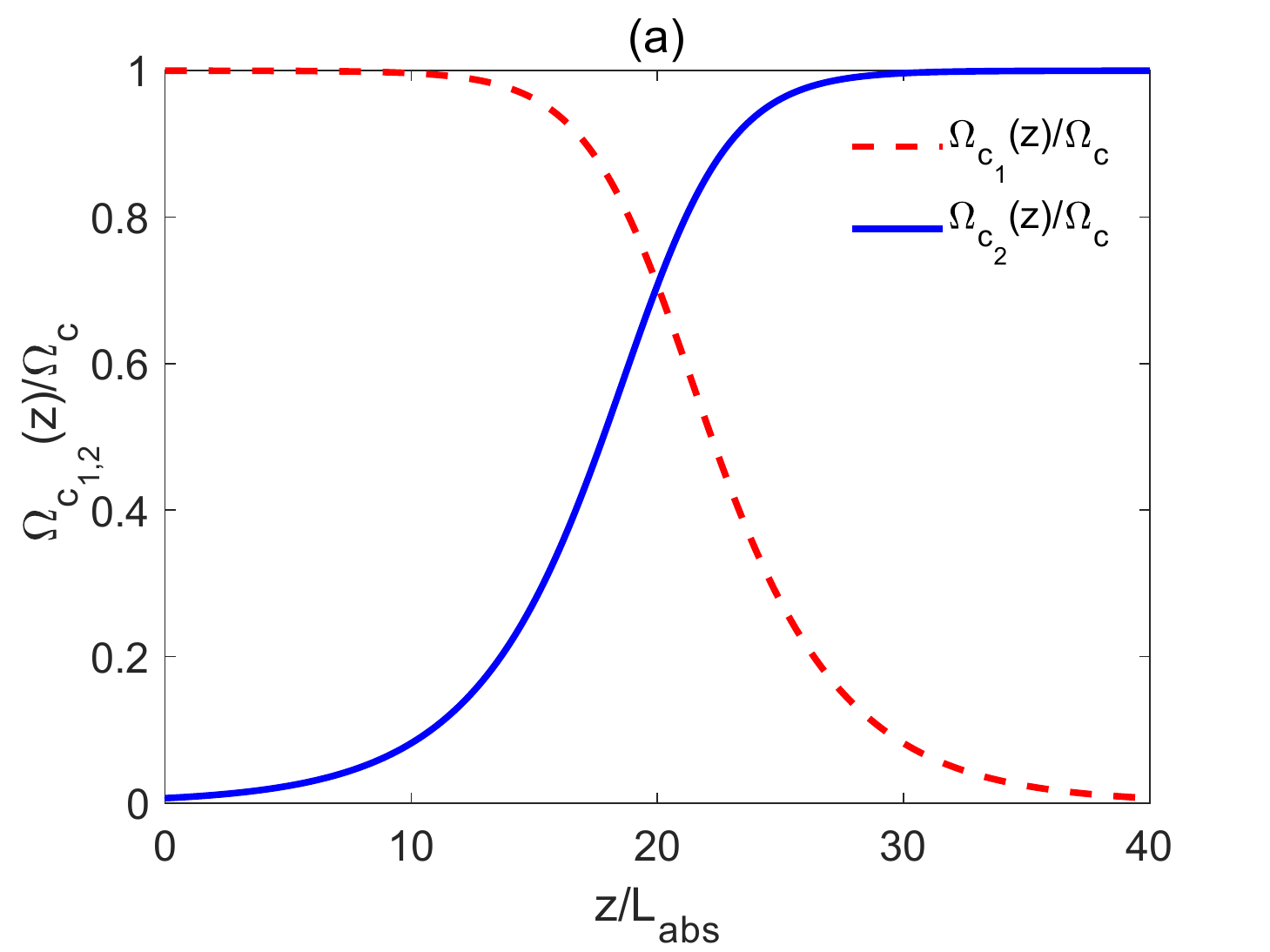} \includegraphics[width=0.4\columnwidth]{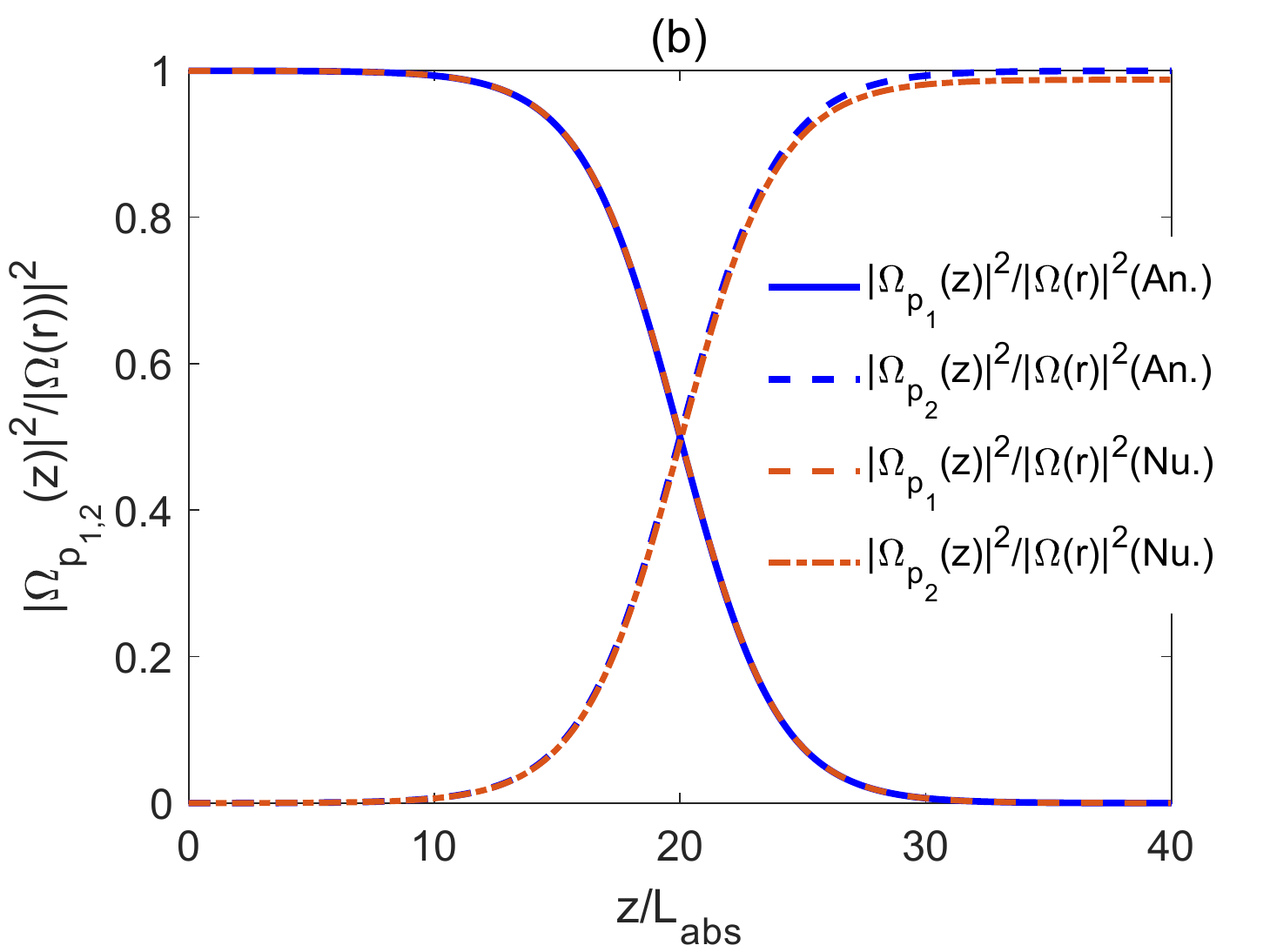}
\caption{(a) Dependence of the dimensionless spatially dependent control fields
given in Eq.~(\ref{eq:Omegac}). (b) Analytical and numerical results
of the dimensionless intensities of the light fields $|\Omega_{p_{1}}(z)|^{2}/|\Omega(r)|^{2}$
and $|\Omega_{p_{2}}(z)|^{2}/|\Omega(r)|^{2}$ against the dimensionless
distance $z/L_{abs}$ and for $\bar{z}=2L_{abs}$, $z_{0}=20L_{abs}$,
$\alpha=40$ and $\delta=0$. Analytical results are plotted based
on Eq.~(\ref{eq:lastSolutions}), while numerical results are based
on the full Optical-Bloch equations given by Eqs. (\ref{eq:A1})-(\ref{eq:A9})
together with the Maxwell equation (\ref{eq:MaxwelEquation}). For
the numerical simulations, the incident form of probe pulses are the
same as Fig.~\ref{fig:2}. Abbreviations \textquotedbl An.\textquotedbl{}
and \textquotedbl Nu.\textquotedbl{} in plots stand for Analytical
and Numerical results. }
\label{fig:3}
\end{figure}

Let us consider a situation where at the beginning of the medium the
first control field is present ($|\Omega_{c_{1}}(z_{i})|=1$) while
the second control field is zero ($|\Omega_{c_{2}}(z_{i})|=0$), giving

\begin{equation}
\left[\begin{array}{c}
|\sin\theta(z_{i})|\\
|\cos\theta(z_{i})|
\end{array}\right]=\left[\begin{array}{c}
1\\
0
\end{array}\right].\label{eq:PDCi}
\end{equation}
Moreover, we assume that the first control field is zero at the end
of the medium $|\Omega_{c_{1}}(z_{f})|=0$ while we have $|\Omega_{c_{2}}(z_{f})|=1$,
resulting 

\begin{equation}
\left[\begin{array}{c}
|\sin\theta(z_{f})|\\
|\cos\theta(z_{f})|
\end{array}\right]=\left[\begin{array}{c}
0\\
1
\end{array}\right],\label{eq:PDCf}
\end{equation}
In that case the efficiency of vortex conversion between different
frequencies is obtained to be the unity using Eqs.~(\ref{eq:lastSolutions})-(\ref{eq:PDCf}): 

\begin{equation}
\eta=|\Omega_{p_{2}}(z_{f})/\Omega(r)|=1,\label{eq:frequency conversion-z}
\end{equation}
This is illustrated in Fig.~\ref{fig:3} showing the dependence of
the intensities $|\Omega_{p_{1}}(z)|^{2}/|\Omega(r)|^{2}$ and $|\Omega_{p_{2}}(z)|^{2}/|\Omega(r)|^{2}$
given by Eq.~(\ref{eq:lastSolutions}) on the dimensionless distance
$z/L_{abs}$ for the resonance case $\delta=0$ and for the optical
depth $\alpha=40$. 

The amplitudes of the spatially dependent control fields satisfying
Eqs.~(\ref{eq:PDCi})-(\ref{eq:PDCf}) can be chosen as 

\begin{equation}
\left[\begin{array}{c}
\Omega_{c_{1}}(z)\\
\Omega_{c_{2}}(z)
\end{array}\right]=\Omega_{c}\left[\begin{array}{c}
\sqrt{\frac{1}{1+e^{(z-z_{0})/\bar{z}}}}\\
\sqrt{\frac{1}{1+e^{-(z-z_{0})/\bar{z}}}}
\end{array}\right],\label{eq:Omegac}
\end{equation}
where $\Omega_{c_{1}}^{2}(z)+\Omega_{c_{1}}^{2}(z)=\Omega_{c}^{2}$
(see Fig.~\ref{fig:3}(a)). Figure~\ref{fig:3}(b) illustrates that
although initially at the beginning of the atomic medium the first
probe field $\Omega_{p_{1}}(z)$ exists, going deeper through the
medium it disappears while the second probe field $\Omega_{p_{2}}(z)$
is generated. 

Note that although Eq.~(\ref{eq:lastSolutions}) does not show to
have any limitations, we have used the adiabatic approximation to
obtain it. This approximation is not valid when the light fields change
too fast in space (or in time). Hence, one needs to find an optimal
condition, when the adiabatic approximation is valid. Inserting the
spatially dependent control fields given by Eq.~(\ref{eq:Omegac})
into the adiabatic condition~(\ref{eq:adiabaticity}) one gets for
the resonant case $\delta=0$:

\begin{equation}
|\frac{1}{\bar{z}}\sqrt{\frac{1}{2+2\cosh\left((z-z_{0})/\bar{z}\right)}}|\ll|\frac{1}{L_{\mathrm{abs}}}|.\label{eq:adiabaticity2}
\end{equation}
At the position where the left hand side of Eq.~(\ref{eq:adiabaticity2})
is the largest ($z=z_{0}$), the condition~(\ref{eq:adiabaticity2})
requires that 

\begin{equation}
\bar{z}\gg L_{\mathrm{abs}}\,.\label{eq:requirment}
\end{equation}
To satisfy the adiabaticity requirement given by Eq.~(\ref{eq:requirment}),
we can take, for example $\bar{z}=2L_{abs}$. The position $z_{0}$
should be in the middle of the atomic medium $z_{0}=\alpha L_{\mathrm{abs}}/2$.
Furthermore to satisfy Eqs.~(\ref{eq:PDCi})\textendash (\ref{eq:PDCf}),
the length of the medium should be much larger than $\bar{z}$. Thus
we can take $z_{0}=20L_{\mathrm{abs}}$ and thus the total length
$40L_{\mathrm{abs}}$.

\begin{figure}
\includegraphics[width=0.4\columnwidth]{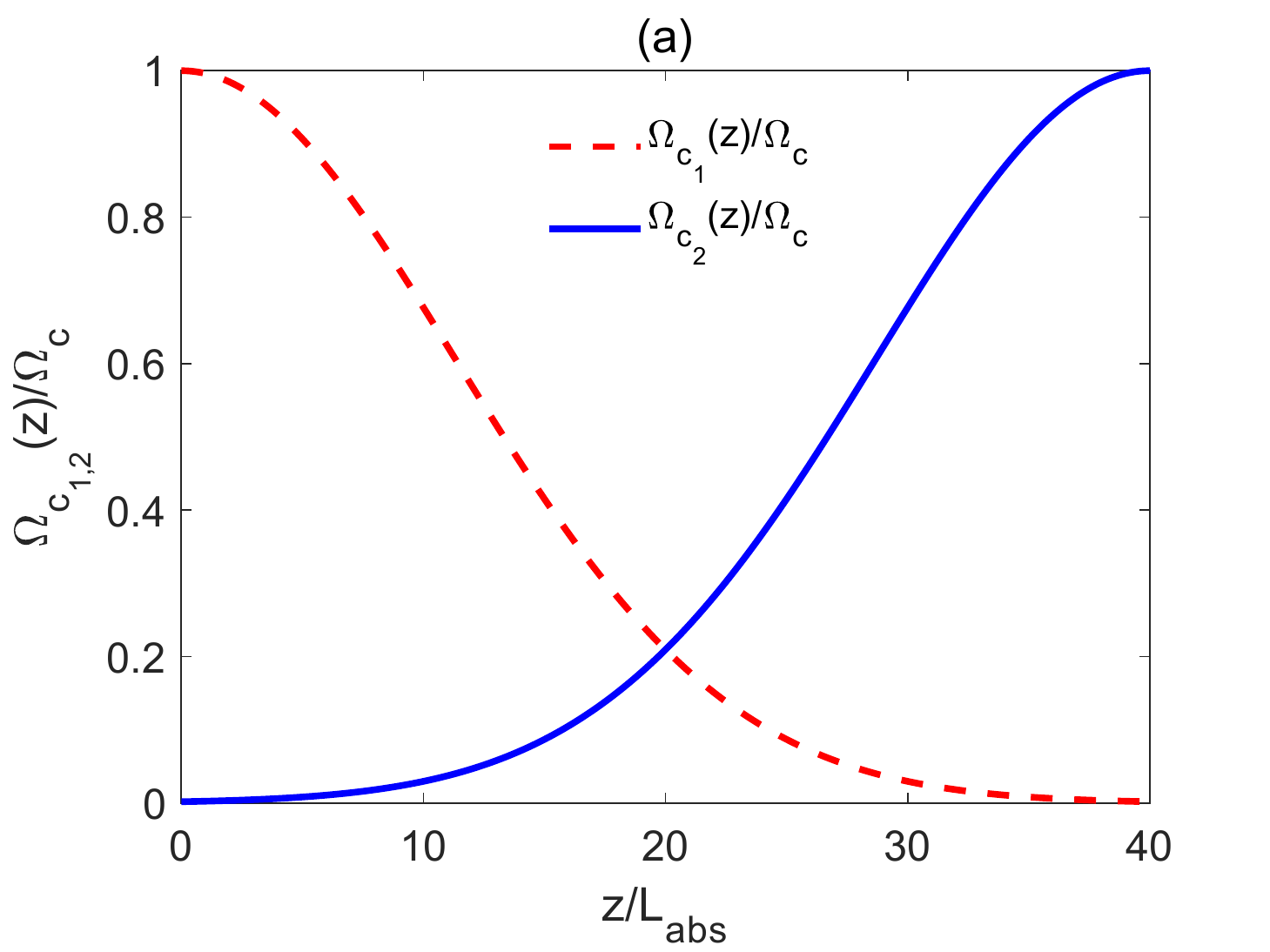} \includegraphics[width=0.4\columnwidth]{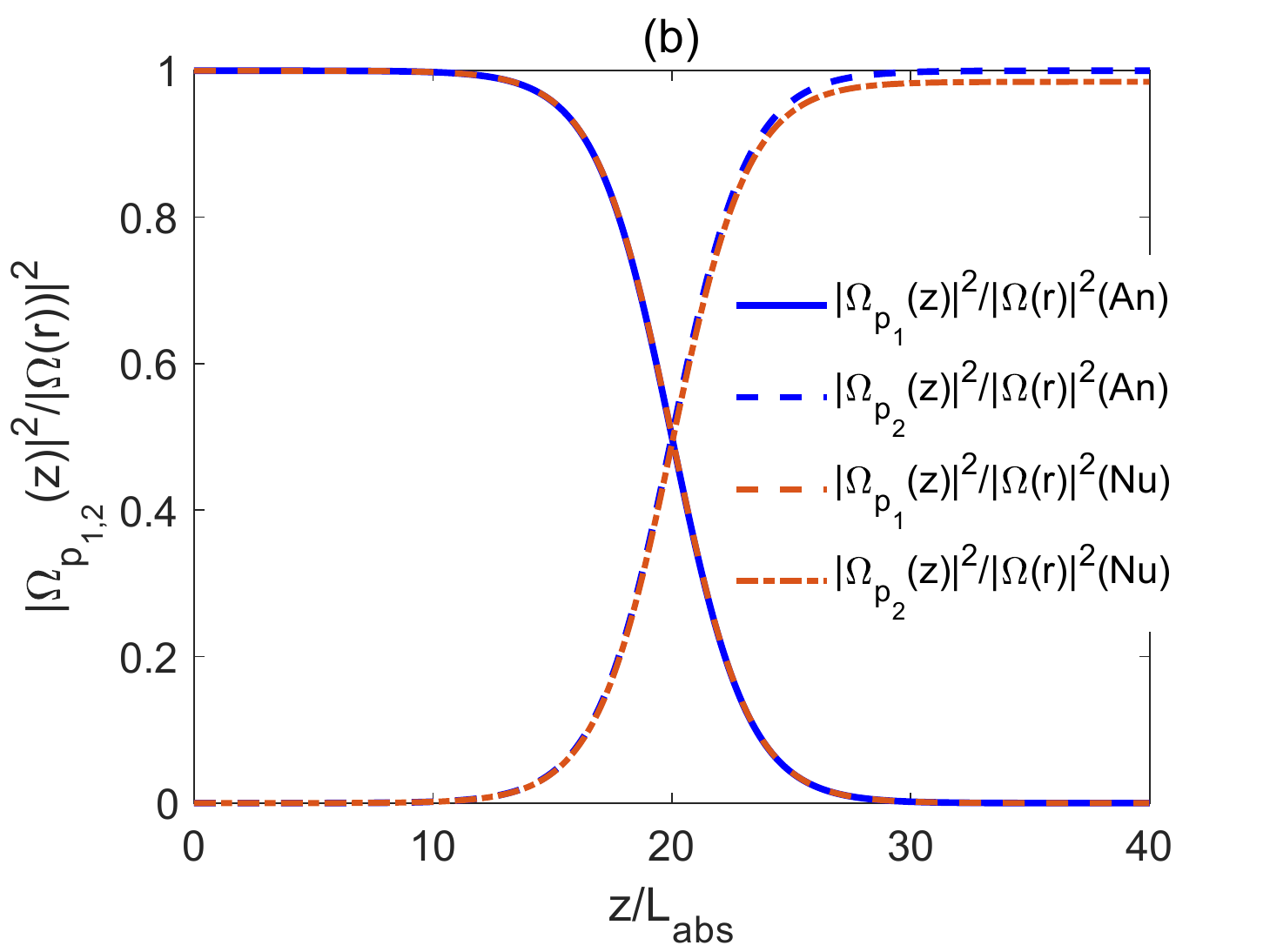}
\caption{(a) Dependence of the dimensionless spatially dependent control fields
given in Eq.~(\ref{eq:Omegac-gaussian}). (b) Analytical and numerical
results of the dimensionless intensities of the light fields $|\Omega_{p_{1}}(z)|^{2}/|\Omega(r)|^{2}$
and $|\Omega_{p_{2}}(z)|^{2}/|\Omega(r)|^{2}$ against the dimensionless
distance $z/L_{abs}$ and for $\alpha=40$, $\sigma=16$ and $\delta=0$.
Analytical results are plotted based on Eq.~(\ref{eq:lastSolutions}),
while numerical results are based on the full Optical-Bloch equations
given by Eqs. (\ref{eq:A1})-(\ref{eq:A9}) together with the Maxwell
equation (\ref{eq:MaxwelEquation}). For the numerical simulations,
the incident form of probe pulses are the same as Fig.~\ref{fig:2}.
Abbreviations \textquotedbl An.\textquotedbl{} and \textquotedbl Nu.\textquotedbl{}
in plots stand for Analytical and Numerical results. }

\label{fig:4}
\end{figure}

Next we work with the spatially dependent Gaussian-shaped control
fields (Fig.~\ref{fig:4}(a)) centered on $z=0$ and $z=L$ and satisfying
Eqs.~(\ref{eq:PDCi})\textendash (\ref{eq:PDCf}) 

\begin{equation}
\left[\begin{array}{c}
\Omega_{c_{1}}(z)\\
\Omega_{c_{2}}(z)
\end{array}\right]=\Omega_{c}\left[\begin{array}{c}
e^{-z^{2}/\sigma^{2}}\\
e^{-(z-L)^{2}/\sigma^{2}}
\end{array}\right],\label{eq:Omegac-gaussian}
\end{equation}
which is experimentally more feasible to produce. Substituting the
Gaussian-shaped control fields given by Eq.~(\ref{eq:Omegac-gaussian})
into the adiabatic condition~(\ref{eq:adiabaticity}) results in
($\delta=0$):

\begin{equation}
|\frac{2Le^{L(L+2z)/\sigma^{2}}}{\sigma^{2}\left(e^{2L^{2}/\sigma^{2}}+e^{4Lz/\sigma^{2}}\right)}|\ll|\frac{1}{L_{\mathrm{abs}}}|.\label{eq:adiabaticity3}
\end{equation}
At the position $z=L/2$ where the left hand side of Eq.~(\ref{eq:adiabaticity3})
has a maximum, one gets

\begin{equation}
\sigma\gg L\sqrt{\frac{1}{\alpha}}.\label{eq:requirment2}
\end{equation}
On the other hand, we want to have intensity on one side of the medium
large, on another side almost zero. This requires

\begin{equation}
\sigma\ll L.\label{eq:requirment3}
\end{equation}
The requirements Eq.~(\ref{eq:requirment2}) and (\ref{eq:requirment3})
together lead to 

\begin{equation}
\alpha\gg1.\label{eq:alphaReq}
\end{equation}
As can be seen from~\ref{fig:4}(b), the first probe field $\Omega_{p_{1}}(z)$
vanishes during propagating through the medium and the second probe
field $\Omega_{p_{2}}(z)$ is generated. For the simulations we select
$\alpha=40$ and $\sigma=16$ satisfying Eqs.~(\ref{eq:requirment2})-(\ref{eq:alphaReq}).

The validity of our adiabatic method is tested by the numerical results
based on the full Maxwell-Bloch equations given by Eqs.~(\ref{eq:A1})-(\ref{eq:A9})
and (\ref{eq:MaxwelEquation}), and a very good agreement is observed,
see Figs.~\ref{fig:3} and \ref{fig:4}. 

\section{Concluding Remarks\label{sec:concl}}

In summary, we have considered propagation of optical vortices in
a cloud of cold atoms characterized by the double-$\Lambda$ configuration
of the atom-light coupling and derived an approximate adiabatic equation~(\ref{eq:lastSolutions})
describing such a propagation. The equation shows that when the adiabaticity
condition (\ref{eq:adiabaticity}) is satisfied, the OAM can be exchanged
between probe pulse pair with a maximum efficiency of $100\%$. It
has been recently shown that a double-$\Lambda$ scheme can be employed
for the exchange of optical vortices based on the EIT \citep{Hamedi-PhysRevA-2018}.
Moreover, the transfer of optical vortices in coherently prepared
$n+1$-level structures has been also demonstrated \citep{Hamedi2019}.
However, a complete energy conversion between optical vortices is
not possible in both previous studies. 

The current approach for the exchange and complete conversion of optical
vortices resembles the STIRAP. Yet in our proposal the roles of the
atomic states and the light fields are interchanged. It is the atomic
states in the STIRAP which are transferred by properly choosing the
time dependence of the light fields. In contrast, in the present situation
the energy is transferred between the light beams by properly choosing
the position dependence of the atomic states. 

The proposed double-$\Lambda$ setup can be implemented experimentally
for example using the $^{87}Rb$ atoms. To have a well-defined polarization,
the control beam propagating orthogonal to the probe should be linearly
polarized along the probe beam direction. The lower levels $|g_{1}\rangle$
and $|g_{2}\rangle$ can then correspond to the $|5S_{1/2},F=1,m_{F}=1\rangle$
$|5S_{1/2},F=2,m_{F}=2\rangle$ hyperfine states. The excited states
$|e_{1}\rangle$ and $|e_{2}\rangle$ can be $|e_{1}\rangle=|5P_{1/2},F=2,m_{F}=2\rangle$
and $|e_{2}\rangle=|5P_{3/2},F=2,m_{F}=2\rangle$, respectively. The
excited state decay rate $\Gamma$ is $\sim2\pi\times6\,MHz$ for
$^{87}Rb$ atoms. We have assumed an optical depth of $40$ which
is experimentally feasible \citep{ChiuPhysRevA2014,Hsiaopra2014}.
Rabi frequencies of the control beams are of the order of $\Gamma$.
In the present study the Rabi frequencies of the probe fields should
be much smaller than the Rabi frequencies of the control beams.
\begin{acknowledgments}
 H.R.H. gratefully thanks professors Thomas Halfmann and Thorsten
Peters for helpful discussions. 
\end{acknowledgments}

\appendix

\section{Full Optical-Bloch equations \label{sec:appendix}}

To make the numerical calculations, we solve the following equations
that describe the time evolution of density matrix operator of the
atoms together with the Maxwell equations for the propagation of probe
fields:

\begin{align}
\dot{\rho}_{e_{2}g_{1}} & =i\Omega_{p_{1}}\rho_{g_{1}g_{1}}+i\Omega_{p_{2}}\rho_{g_{2}g_{1}}-i\Omega_{c_{1}}\rho_{e_{2}e_{1}}-i\Omega_{p_{1}}\rho_{e_{2}e_{2}}-(\Gamma_{e_{2}g_{1}}+i\Delta_{e_{2}g_{1}})\rho_{e_{2}g_{1}},\label{eq:A1}\\
\dot{\rho}_{e_{2}g_{2}} & =i\Omega_{p_{2}}\rho_{g_{2}g_{2}}+i\Omega_{p_{1}}\rho_{g_{1}g_{2}}-i\Omega_{c_{2}}\rho_{e_{2}e_{1}}-i\Omega_{p_{2}}\rho_{e_{2}e_{2}}-(\Gamma_{e_{2}g_{2}}+i\Delta_{e_{2}g_{2}})\rho_{e_{2}g_{2}},\label{eq:A2}\\
\dot{\rho}_{e_{2}e_{1}} & =i\Omega_{p_{1}}\rho_{g_{1}e_{1}}+i\Omega_{p_{2}}\rho_{g_{2}e_{1}}-i\Omega_{c_{1}}\rho_{e_{2}g_{1}}-i\Omega_{c_{2}}\rho_{e_{2}g_{2}}-(\Gamma_{e_{2}e_{1}}+i\Delta_{e_{2}e_{1}})\rho_{e_{2}e_{1}},\label{eq:A3}\\
\dot{\rho}_{e_{1}g_{1}} & =i\Omega_{c_{1}}\rho_{g_{1}g_{1}}+i\Omega_{c_{2}}\rho_{g_{2}g_{1}}-i\Omega_{c_{1}}\rho_{e_{1}e_{1}}-i\Omega_{p_{1}}\rho_{e_{1}e_{2}}-(\Gamma_{e_{1}g_{1}}+i\Delta_{e_{1}g_{1}})\rho_{e_{1}g_{1}},\label{eq:A4}\\
\dot{\rho}_{e_{1}g_{2}} & =i\Omega_{c_{2}}\rho_{g_{2}g_{2}}+i\Omega_{c_{1}}\rho_{g_{1}g_{2}}-i\Omega_{c_{2}}\rho_{e_{1}e_{1}}-i\Omega_{p_{2}}\rho_{e_{1}e_{2}}-(\Gamma_{e_{1}g_{2}}+i\Delta_{e_{1}g_{2}})\rho_{e_{1}g_{2}},\label{eq:A5}\\
\dot{\rho}_{g_{2}g_{1}} & =i\Omega_{c_{2}}\rho_{e_{1}g_{1}}+i\Omega_{p_{2}}\rho_{e_{2}g_{1}}-i\Omega_{c_{1}}\rho_{g_{2}e_{1}}-i\Omega_{p_{1}}\rho_{g_{2}e_{2}}-(\Gamma_{g_{2}g_{1}}+i\Delta_{g_{2}g_{1}})\rho_{g_{2}g_{1}},\label{eq:A6}\\
\dot{\rho}_{g_{1}g_{1}} & =i\Omega_{c_{1}}(\rho_{e_{1}g_{1}}-\rho_{g_{1}e_{1}})+i\Omega_{p_{1}}(\rho_{e_{2}g_{1}}-\rho_{g_{1}e_{2}})+\gamma_{e_{1}g_{1}}\rho_{e_{1}e_{1}}+\gamma_{e_{2}g_{1}}\rho_{e_{2}e_{2}},\label{eq:A7}\\
\dot{\rho}_{g_{2}g_{2}} & =i\Omega_{c_{2}}(\rho_{e_{1}g_{2}}-\rho_{g_{2}e_{1}})+i\Omega_{p_{2}}(\rho_{e_{2}g_{2}}-\rho_{g_{2}e_{2}})+\gamma_{e_{1}g_{2}}\rho_{e_{1}e_{1}}+\gamma_{e_{2}g_{2}}\rho_{e_{2}e_{2}},\label{eq:A8}\\
\dot{\rho}_{e_{2}e_{2}} & =i\Omega_{p_{1}}(\rho_{g_{1}e_{2}}-\rho_{e_{2}g_{1}})+i\Omega_{p_{2}}(\rho_{g_{2}e_{2}}-\rho_{e_{2}g_{2}})-(\gamma_{e_{2}g_{1}}+\gamma_{e_{2}g_{2}})\rho_{e_{2}e_{2}},\label{eq:A9}
\end{align}
where $\rho_{g_{1}g_{1}}+\rho_{g_{2}g_{2}}+\rho_{e_{1}e_{1}}+\rho_{e_{2}e_{2}}=1$.
In numerical simulations we take $\gamma_{e_{1}g_{1}}=\gamma_{e_{2}g_{2}}=\gamma_{e_{1}g_{2}}=\gamma_{e_{2}g_{1}}=\Gamma$,
$\Gamma_{e_{2}g_{1}}=\Gamma_{e_{2}g_{2}}=\Gamma_{e_{1}g_{1}}=\Gamma_{e_{1}g_{2}}=\Gamma$,
$\Gamma_{g_{1}g_{2}}=0$, $\Gamma_{e_{2}e_{1}}=2\Gamma$, $\Delta_{e_{2}g_{1}}=\Delta_{e_{2}g_{2}}=\delta$
and $\Delta_{e_{2}e_{1}}=\Delta_{e_{1}g_{1}}=\Delta_{e_{1}g_{2}}=\Delta_{g_{2}g_{1}}=0$.

\end{document}